\documentclass[aps,prb,twocolumn]{revtex4}
\usepackage{graphicx}
\usepackage{url}
\usepackage{natbib}

\begin{document}
\title{Absolute values of the London penetration depth in YBa$_2$Cu$_3$O$_{6+{\rm y}}$ measured by zero field
ESR spectroscopy on Gd doped single crystals}

\author{T.~Pereg-Barnea}
\author{P.~J.~Turner}
\author{R.~Harris}
\author{G.~K.~Mullins}
\author{J.~S.~Bobowski}
\author{M.~Raudsepp}
\altaffiliation{Department of Earth and Ocean Sciences,
University of British Columbia, 6339 Stores Road, Vancouver, British Columbia, Canada V6T 1Z4}
\author{Ruixing~Liang}
\author{D.~A.~Bonn}
\author{W.~N.~Hardy}
\affiliation{Department of Physics and Astronomy,
University of British Columbia, 6224 Agricultural Road, Vancouver, British Columbia, Canada V6T 1Z1}
\date{\today}

\begin{abstract}
Zero-field electron spin resonance (ESR) of dilute Gd ions substituted for Y in the
cuprate superconductor YBa$_2$Cu$_3$O$_{\rm 6+y}$ is used as a novel technique for
measuring the absolute value of the low temperature magnetic penetration depth
$\lambda(T\to 0)$. The Gd ESR spectrum of samples with $\approx 1\%$ substitution was
obtained with a broadband microwave technique that measures power absorption
bolometrically from 0.5~GHz to 21~GHz. This ESR spectrum is determined by the crystal
field that lifts the level degeneracy of the spin $7/2$ Gd$^{3+}$ ion and details of this
spectrum provide information concerning oxygen ordering in the samples. The magnetic
penetration depth is obtained by relating the number of Gd ions exposed to the microwave
magnetic field to the frequency-integrated intensity of the observed ESR transitions.
This technique has allowed us to determine precise values of $\lambda$ for screening
currents flowing in the three crystallographic orientations ($\hat a$, $\hat b$ and $\hat
c$) in samples of Gd$_{\rm x}$Y$_{\rm 1-x}$Ba$_2$Cu$_3$O$_{6+{\rm y}}$ of three different
oxygen contents ${\rm y}=0.993$ ($T_c = 89$~K), ${\rm y}=0.77$ ($T_c=75$~K) and ${\rm
y}=0.52$ ($T_c=56$~K). The in-plane values are found to depart substantially from the
widely reported relation $T_c\propto 1/\lambda^2$.
\end{abstract}

\maketitle
\section{Introduction}

The magnetic penetration depth $\lambda$ figures prominently in many aspects of
superconductivity. It is the length scale over which an external magnetic field is
screened by a superconductor in the Meissner state, it sets the size of vortices in the
mixed state, and it is a controlling factor in many other properties such as microwave
and far infrared absorption. The fundamental importance of the penetration depth is that
it provides a rather direct measure of the superfluid phase stiffness $\rho_s \propto
1/\lambda^2$, or what is often referred to as superfluid density $n_s/m^* =
1/\mu_0e^2\lambda^2$, which can be a tensor quantity due to effects such as an
anisotropic effective mass $m^*$. In the cuprates, the dependence of $\lambda$ on
temperature $T$ and hole-doping have provided major pieces of the cuprate puzzle.
Microwave measurements of $\Delta\lambda(T)$ at low $T$ revealed a linear dependence
providing some of the first key evidence of the presence of line nodes in the
superconducting gap in the cuprates\cite{Walter} and the temperature dependence of
$\lambda(T)$ near $T_c$ suggested that the transition is governed by 3DXY critical
fluctuations.\cite{Saeid3dxy} Evidence that the low temperature value of the phase
stiffness $\rho_s(T\to 0)$
depends linearly on $T_c$, has been obtained
from muon spin relaxation measurements of $\lambda(T)$.\cite{Uemura,Sonier1}

Accurate and precise measurements of the absolute value of $\lambda(T\rightarrow 0)$ are
particularly important for understanding superconductivity in the cuprates, and are
critical for broad studies of doping dependence. Having an absolute value of $\lambda$ is
also crucial in the interpretation of other experiments that involve superfluid
screening, with examples including the microwave conductivity, the lower critical field
$H_{C1}$, and the conversion of measurements of $\Delta\lambda(T)$ to the temperature
dependent superfluid density $\propto 1/\lambda^2(T)$. However, the absolute value of
$\lambda$ has proven to be particularly difficult to measure. A wide variety of
techniques exist for doing so, but each has important drawbacks that must be
acknowledged. One class of techniques takes place in the mixed state, probing the length
scale over which fields decay away from a magnetic vortex.\cite{Sonier1} Muon spin
relaxation ($\mu SR$) falls into this class and suffers from the difficulty that the
screening length scale obtained from a mixed state measurement in relatively high fields
can differ from the penetration depth in the Meissner state because of non-linear and
non-local effects.\cite{Sonier1} Vortex state measurements are also unable to directly
resolve the anisotropy in $\lambda$ in orthorhombic materials because the screening
currents circulate around a vortex. A second class of techniques measures magnetic flux
exclusion from a sample in the Meissner state, essentially by measuring the magnetic
susceptibility at DC, radio or microwave frequencies.\cite{Krusin,Carrington,Walter}
These can precisely determine the temperature dependence $\Delta\lambda(T) =
\lambda(T)-\lambda(T_o)$ relative to a base temperature $T_o$. However, a measurement of
the absolute value amounts to a comparison between the physical volume of the sample and
a susceptibility determination of the volume that is field-free, the difference being the
small volume at the surface that is penetrated by the field. Such a measurement is most
sensitive if the ratio of surface area to volume is large, which can be achieved by
working with powders. However, difficulties in aligning powders and modelling the shape
and size distribution of grains introduce substantial uncertainties in the absolute value
of $\lambda$.\cite{Panagopoulos} For a macroscopic, high quality crystal of a cuprate
superconductor, it is nearly impossible to measure sample dimensions precisely enough to
achieve an absolute measurement of $\lambda$ this way, except in geometries dominated by
the very large values of $\lambda$ for screening currents running in the
$\hat{c}$-direction.

A very direct measurement of the penetration depth has been achieved by measuring the
transmission of low frequency fields through very thin films using mutual inductance
techniques. While this is a very sensitive method, its chief drawbacks are the need to
work with films and the difficulty of measuring anisotropy.\cite{Lemberger1} Techniques
that can resolve the anisotropy include far infrared and optical measurements that
determine the inductive response of the superfluid at high frequencies.\cite{Homes,Basov}
Such measurements are difficult because of their need for spectra over a very wide
frequency range in order to perform Kramers-Kr\"{o}nig transforms and are also
susceptible to error because the measured response can also include screening by
uncondensed carriers in addition to the superfluid. However, this approach can be used to
determine the anisotropy of $\lambda$ simply by polarizing the light and measuring
optical properties along all of the principal axes. Such infrared determinations of the
absolute value of $\lambda$ at low temperature have been combined with microwave
measurements of $\Delta\lambda(T)$ to produce the anisotropic $1/\lambda^2(T)$ in
YBa$_2$Cu$_3$O$_{6+{\rm y}}$ at y=0.60 and 0.95,\cite{BonnCzech} but it would be far
preferable to have a more direct microwave measurement that gives the absolute value of
$\lambda$, measured {\it on the same sample} as that used to determine the temperature
dependence.

This paper describes a microwave technique that directly measures the small volume near
the surface that is field-penetrated in the Meissner state, rather than inferring
$\lambda$ from a measurement of the very large volume that is field-free (see
Fig.~\ref{fig:fields}). We use this technique here to measure the absolute value of
$\lambda$ for all three principal axes in YBa$_2$Cu$_3$O$_{6+{\rm y}}$. The key to the
method is to embed a randomly distributed low concentration of magnetic moments in the
superconductor in such a way that they will act as local field probes and yet will
minimally affect the transport properties. The magnetic moments will serve as a
non-interacting spin system whose electron spin resonance (ESR) spectrum is a sum of
one-particle ESR transitions. The energy levels of a single spin are determined by its
crystallographic environment, {\em i.e.} the charge distribution around the magnetic ion,
and the resulting susceptibility is that of a single spin multiplied by the number of
spins that participate in the ESR process. If the moments are evenly distributed in the
sample, the number of spins that participate in the process is proportional to the fraction
of the sample's volume that is exposed to the field, {\em i.e.} the volume within a
penetration depth of the surface. In order to count the number of spins exposed, we
measure the zero field ESR absorption spectrum with a microwave frequency magnetic field
applied perpendicular to the principal spin axis acting as a small perturbation.

In the next section of this paper we describe the samples, the broadband microwave
apparatus and the measurements of the apparent microwave surface resistance that
ultimately yield values of the penetration depth. The third section describes how the
theoretical ESR spectrum was calculated using an effective crystal field Hamiltonian,
which is a necessary prerequisite for extracting $\lambda$. In the fourth section we
examine the issue of oxygen ordering in the CuO chains of YBa$_2$Cu$_3$O$_{6+{\rm y}}$
and comment upon the sensitivity of our measurements to this ordering. Finally, we
summarize our results and present new values for $\lambda$ in all three directions, for
three different dopings in the YBa$_2$Cu$_3$O$_{6+{\rm y}}$ system. These results
indicate a substantial departure from $T_c\propto 1/\lambda^2$, the so-called Uemura
scaling.

\section{Sample Preparation and Experimental Techniques}
\begin{figure}
\includegraphics[width=\columnwidth]{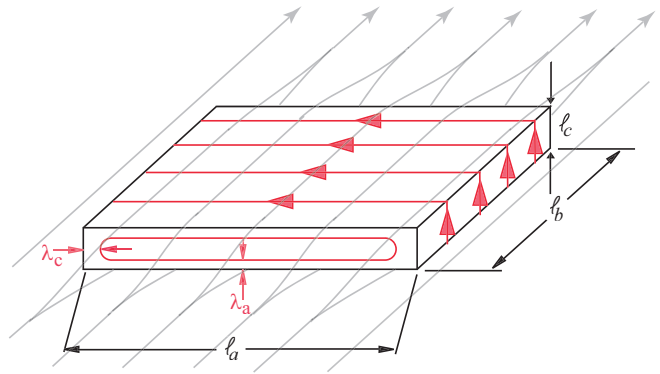}
\caption{\label{fig:fields} Schematic drawing of a platelet sample of a superconductor
having an anisotropic penetration depth $\lambda$ in a uniform applied microwave magnetic
field $H_{r\! f}$. In the above picture, the field penetrates a volume determined by
$\lambda_a$ and $\lambda_c$ and the crystal dimensions $\ell_a$ and $\ell_c$, but
contains no contribution from $\lambda_b$. By combining measurements where we rotate and
cleave the crystal, we are able to determine values of $\lambda$ for all three
crystallographic directions.}
\end{figure}

The objective of this work is to provide accurate measurements of
the intrinsic absolute value of the magnetic penetration depth in
the limit $T \to 0$. It is well-known that impurity doping can
have a strong influence on the low energy density of states in a
$d$-wave superconductor,\cite{hirschfeld-atkinson,bonnZn} and for
the present work care must be taken to ensure that the
introduction of the spin-probe impurity does not alter
$\lambda(T\to0)$. This constrains the Gd concentration x, but
conversely we wish to maximize the amplitude of the ESR response,
which scales as x. We have found a good compromise at a nominal
value of x$\approx 1\%$ which results in an easily resolved ESR
spectrum and does not significantly alter the intrinsic low temperature
properties of the material.

In Gd$_{\rm x}$Y$_{\rm 1-x}$Ba$_2$Cu$_3$O$_{6+{\rm y}}$, the dopant Gd$^{3+}$ ion
substitutes for the Y$^{3+}$ ion which is sandwiched between the two CuO$_2$ planes in
the YBa$_2$Cu$_3$O$_{6+{\rm y}}$ unit cell that support the bulk of the electronic
transport. The energy cost for cross-substitution of the Gd on the Ba site is prohibitive
and such cation substitution occurs only for the lighter and larger members of the
rare earth series. The ESR spectrum is determined completely by the crystal field
environment of the Gd$^{3+}$ ion and therefore knowing that there is only one site for
substitution means that changes in the ESR spectrum from sample-to-sample can only be due
to changes of the oxygen order in the CuO chains. Another important feature of
substitution on the Y site is that the Gd impurities are expected to be weak (Born) limit
quasiparticle scattering centers as was found to be the case for Ca$^{2+}$ substitution
on the same site.\cite{BonnCzech} Born scatterers have little effect on the superfluid
density in the low temperature limit. This is in contrast, for instance, to the case of
the non-magnetic impurity Zn$^{2+}$ which substitutes {\em into} the CuO$_2$ plane, and
is known to be strongly pair-breaking.

The Gd$_{\rm x}$Y$_{\rm 1-x}$Ba$_2$Cu$_3$O$_{6+{\rm y}}$ samples were grown using a
self-flux method in BaZrO$_3$ crucibles as described elsewhere.\cite{Ruixing} The
starting materials had at least 99.999~at.~\% purity with the nominal concentration of Gd
ions set by the ratio of oxide precursor substitution Gd$_2$O$_3$:Y$_2$O$_3$. The
as-grown platelet single crystals were mechanically detwinned, annealed in flowing oxygen
to set the oxygen content, and given a final anneal at lower temperature for oxygen
ordering. The annealing parameters have been previously published for the slightly
overdoped (nearly full CuO chains, y=0.993)\cite{Ruixing} and ortho-II ordered
(alternating full and empty chains, y=0.52)\cite{RuixingII} crystals. Other ordered
phases at intermediate doping can also be produced with larger chain-ordered
superlattices, namely ortho-III\cite{Manca} and ortho-VIII.\cite{Zimmermann} Crystals at
these doping levels were produced using procedures very similar to that used to produce
the ortho-II samples, however the initial temperatures for setting the oxygen content
were different (668 $^\circ$C for ortho-VIII and 609 $^\circ$C for ortho-III) as were the
final low temperature annealing temperatures used to establish oxygen ordering in the CuO
chain layers (35 $^\circ$C for ortho-VIII (y=0.67) order and 60 $^\circ$C for ortho-III
(y=0.77) order).

A critical parameter in our data analysis is the absolute concentration of Gd ions x,
which was measured using electron-probe micro-analysis (EPMA). We used a fully-automated
CAMECA SX-50 instrument, operated in the wavelength-dispersion mode with the following
operating conditions:  excitation voltage, 15~kV; beam current, 20~nA; peak count time
(240~s for Gd), 20~s; background count-time, 10~s each side of peak (120~s for Gd); spot
diameter, 5~$\mu$m.  Data reduction was done using the ``PAP" $\phi$($\rho$Z)
method.\cite{EPMA}  For the elements considered, the following standards, X-ray lines and
crystals were used: Ga$_3$Gd$_5$O$_{12}$, GdL$\alpha$, LiF; YBa$_2$Cu$_3$O$_{6.95}$,
BaL$\alpha$, PET; YBa$_2$Cu$_3$O$_{6.95}$, CuK$\alpha$, LiF; YBa$_2$Cu$_3$O$_{6.95}$,
OK$\alpha$, W/Si multilayer dispersion element. For the present study, we found it to be
important that each Gd$_{\rm x}$Y$_{\rm 1-x}$Ba$_2$Cu$_3$O$_{6+{\rm y}}$ crystal was
examined individually. Because the sample must be embedded in epoxy for EPMA, this was
done following all microwave measurements. For a nominal Gd concentration of x$=1\%$ in
the growth flux, the resulting concentration in the crystals was found vary to over the
range from x$=1.05(.06)\%$ to x$=1.37(.12)\%$. At least 10 locations were measured on
each crystal surface as a means of ensuring that the stoichiometry was homogeneous. In
all but one sample, the uniformity was within the statistical limitations of the
measurement. For one sample it was necessary to examine 54 different locations covering
the crystal surface to map out and take account of a small region, approximately 5$\%$ of
the total surface, having an elevated Gd concentration of x$=1.50(.08)\%$. Measurements
on the cut edges of a 100~$\mu m$ thick sample confirmed that there was no gradient through
the thickness of the sample. If the Gd concentration in the melt were to drift with time
during the crystal growth, then a resulting Gd gradient in the crystal would be
problematic for the following analysis.

The broadband microwave absorption spectrum (0.5~GHz to 21~GHz) was measured using a
novel spectrometer based on a bolometric method of detection, described in detail
elsewhere.\cite{PatRSI} In essence, the single crystal Gd$_{\rm x}$Y$_{\rm
1-x}$Ba$_2$Cu$_3$O$_{6+{\rm y}}$ sample is exposed to a spatially uniform microwave
magnetic field that is amplitude modulated at low frequency. The corresponding
temperature oscillations of the sample are detected synchronously, and absolute
calibration of the thermal stage sensitivity allows us to infer the magnitude of the
power absorbed by the sample. A critical ingredient of the method is a well-characterized
normal-metal reference sample that is mounted on a second thermal stage and placed in a
symmetric location in the microwave transmission line. This allows us to monitor the
strength of the applied microwave field which varies strongly with frequency due to
standing waves in the microwave circuit.

At microwave frequencies in the limit of local electrodynamics,\cite{Ahmad} the
experimentally measurable quantity is the surface impedance Z$_s$ which is related to the
complex conductivity $\sigma=\sigma_1-{\rm i}\sigma_2$ via
\begin{equation}
\label{eqn:Zs} \textrm{Z}_s \equiv \textrm{R}_s+\textrm{iX}_s=\sqrt{
{\textrm{i}\mu_o\omega}\over{\sigma_1-\textrm{i}\sigma_2} }.
\end{equation}
For a superconductor well-below $T_c$ measured at low frequency, the high
superfluid density ensures that the response is mainly reactive ({\em i.e.}
$\sigma_2 \gg \sigma_1$) and Eq.~\ref{eqn:Zs} simplifies to
\begin{eqnarray}
\label{eqn:RsXs}
\textrm{R}_s(\omega,T) &\simeq& {1\over2} \mu_o^2\omega^2\lambda^3(T)\sigma_1(\omega,T), \\
\textrm{X}_s(\omega,T) &\simeq&  \sqrt{ {\mu_o\omega}\over{\sigma_2}}
=\mu_o\omega\lambda(T). \nonumber
\end{eqnarray}
(The appearance of $\lambda$ in both of the above expressions highlights its
important role in determining the conductivity from measurements of $R_s$ and
$X_s$.) The power absorption is determined by the surface resistance R$_s\equiv
Re\mid$Z$_s\!\mid$ according to
\begin{equation}
\label{eqn:power} \ P_{abs}= R_s \int H_{r\! f}^2 \textrm{d}S,
\end{equation}
where $H_{r\! f}$ is the root-mean-square (rms) magnitude of the uniform, tangential
magnetic field at the surface $S$ of the sample. At low temperatures, absorption in a
superconductor is due to quasiparticles thermally excited from the condensate, and the
study of the quasiparticle conductivity spectrum $\sigma_1(\omega,T)$ has been the
central focus of other recent work.\cite{PatPRL,Ahmad} In the present case of magnetic
impurities in a superconductor, the imaginary part of the magnetic susceptibility,
$\chi''$, provides another mechanism for power absorption.  Formally, the magnetic
response can be included by a modification of the vacuum permeability $\mu_o$ in
Eq.~\ref{eqn:Zs} to include a dimensionless complex susceptibility $\chi$, written as
$\mu_o\to\mu=\mu_o (1+\chi'-\textrm{i}\chi'')$. For the case of a sufficiently low
concentration of magnetic impurities, $\chi',\chi''\ll1$, and the apparent surface
impedance expressions can then be rewritten as
\begin{eqnarray}
\label{eqn:RsXsApp}
\textrm{R}_s^{app}(\omega,T) &\simeq& {1\over2} \mu_o^2\omega^2\lambda^3(T)\left[\sigma_1(\omega,T)+ {\chi''(\omega,T) \over \mu_o\omega\lambda^2(T)}\right] ,~~~~ \\
\textrm{X}_s^{app}(\omega,T) &\simeq& \mu_o\omega\lambda(T)
\sqrt{1+\chi'(\omega,T)}. \nonumber
\end{eqnarray}
Thus, a frequency scanned measurement of the power absorption will provide an
apparent surface resistance with two separate contributions: one from the
quasiparticle conductivity $\sigma_1(\omega,T)$ and one from the ESR spectrum
contained in $\chi''(\omega,T)$.

\begin{figure}
\includegraphics[width=\columnwidth]{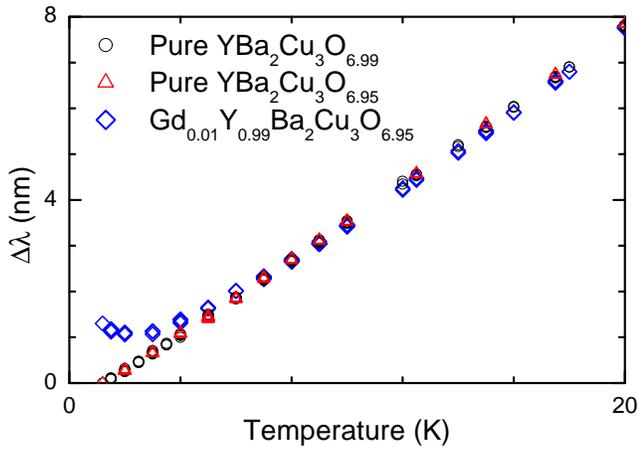}
\caption{\label{fig:lambdaupturn} Measurements of the temperature-dependent
change in apparent magnetic penetration depth, $\Delta\lambda(T)$, performed
via cavity perturbation in a loop-gap resonator operating at 954~MHz. A
comparison between measurements on single crystal samples with and without 1\%
Gd substituted for Y demonstrates that only below a temperature of $\sim$7~K
does the contribution from the real part of the susceptibility
$\chi'(\omega,T)$ contribute to the measurement as an upturn in the apparent
penetration depth. Above this temperature, the intrinsic linear dependence of
$\Delta \lambda(T)$ of a clean $d$-wave superconductor is recovered.}
\end{figure}

\begin{figure}
\includegraphics[width=\columnwidth]{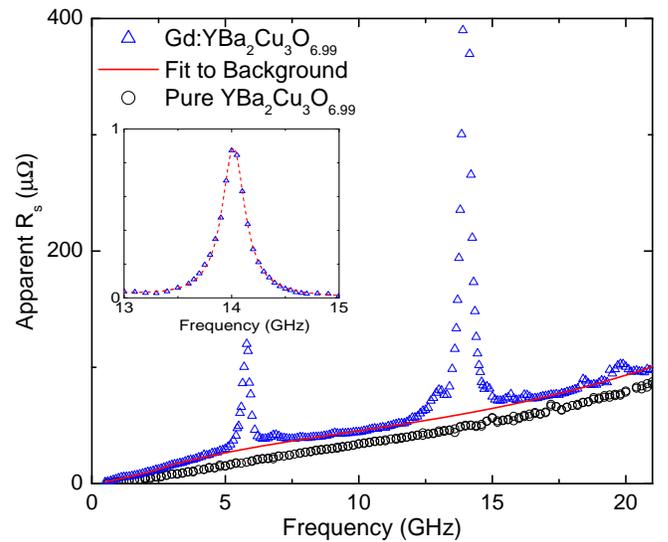}
\caption{\label{fig:RsRaw} Broadband microwave absorption measurement of a
Gd$_{0.01}$Y$_{0.99}$Ba$_2$Cu$_3$O$_{6.99}$ single crystal sample at 1.3~K. The
broad, smooth background is due to quasiparticle absorption and the sharp lines
are due to ESR transitions. The smooth line is a polynomial fit to the
background away from the ESR peaks, and is seen to be similar to a sample
without Gd. The inset shows that the main ESR transition at 14.05~GHz is very
well described by a Lorentzian lineshape, shown as a dashed line. }
\end{figure}

Equation~\ref{eqn:RsXsApp} reveals that a measurement of the surface reactance
will contain a contribution from $\chi'(\omega,T)$ in addition to probing
$\lambda(T)$. This is evident in the comparison of our measurements of $\Delta
\lambda(T)$ in pure and Gd doped YBa$_2$Cu$_3$O$_{6.99}$, presented in
Fig.~\ref{fig:lambdaupturn}. In fact, at the outset of this work we considered
the possibility of fitting the curvature arising from $\chi'(T)$ in a low
frequency measurement of $\Delta \lambda(T)$ as a means of extracting the
absolute value of $\lambda$. However, it was decided that this method was not
practical as it lacked the necessary sensitivity and also required knowing
$\chi'(\omega,T)$ {\em a priori}. A further complication to this approach is
that it has been well-established that a variety of paramagnetic impurities can
generate a low temperature upturn in a low frequency $\Delta \lambda(T)$
measurement,\cite{prozorov} which would produce systematic errors in the
extraction of an absolute value of $\lambda$.

Fortunately, the broadband spectroscopy apparatus provides a means of accessing the ESR
spectrum in  much greater detail. Figure~\ref{fig:RsRaw} shows a measurement of
R$_s^{app}(\omega)$ from 0.5~GHz to 21~GHz at 1.3~Kelvin for a
Gd$_{0.01}$Y$_{0.99}$Ba$_2$Cu$_3$O$_{6.99}$ sample.  The quasiparticle absorption
spectrum is known to be a slowly varying, monotonically increasing function of frequency,
and the sharp ESR absorption lines ($\Delta {\rm f}<$~0.5~GHz) are superposed. We have
shown previously that the low temperature microwave conductivity $\sigma_1(\omega,T)$
extracted from measurements of $R_s(\omega,T)$ can be interpreted as that of weak-limit
quasiparticle scattering from impurities,\cite{PatPRL} and here we attribute the small
changes in the spectrum of Fig.~\ref{fig:RsRaw} to a slight increase in scattering due to
the increased density of out-of-plane defects in the Gd doped samples. In order to
separate the ESR contribution from the quasiparticle conductivity, we fit to the smooth
background away from the ESR peaks using a second order polynomial, shown as a solid line
in Fig.~\ref{fig:RsRaw}. After subtracting the polynomial, the remaining quantity is
equal to ${1\over2} \mu_o\omega\lambda \chi''(\omega)$ (from Eq.~\ref{eqn:RsXsApp}). In
the subsequent section, we discuss the calculation of the quantity $\chi''(\omega)$ from
the effective spin Hamiltonian. With the theoretical curve for the susceptibility of one
Gd spin in hand, we are able to fit to the experimentally determined quantity with an
overall multiplicative fit parameter $\lambda$.

YBa$_2$Cu$_3$O$_{6+{\rm y}}$ single crystals grow naturally as platelets having broad
$\widehat{ab}$-plane crystal faces and thin $\hat{c}$-axis dimensions. All of our
microwave measurements involve a uniform radio frequency magnetic field $H_{r\! f}$
applied parallel to the broad face of the crystal in order to minimize the effects of
demagnetization, as shown schematically in Fig.~\ref{fig:fields}. It is also known that
the crystal $\hat{c}$-axis is the principal spin axis of the Gd ions, so that the ESR
response is maximized in this configuration. The number of Gd spins that are exposed to
the microwave field is governed by the effective penetration depth $\lambda_{e\! f\! f}$
and, for our geometry, will always contain contributions from $\lambda_a$ or $\lambda_b$
and $\lambda_c$. In the cuprates, the two-dimensionality of the CuO$_2$ planes results in
a large anisotropy in the penetration depth: $\lambda_c \gg \lambda_{a},\lambda_{b}$ for
all doping levels. In practice, this means that despite working with thin crystals having
the in-plane dimension $\ell_{a,b} \gtrsim 10\ell_c$, the $\hat{c}$-axis contribution to
$\lambda_{e\! f\! f}$ can be large. In some cases it was also necessary to account for a
small fraction $t$ of the crystal which remained twinned (14\% at most), introducing the
third component of $\lambda$ into the measurement. Since $\lambda$ is small compared to
any crystal dimension, it suffices to use a simple linearized relation for $\lambda_{e\!
f\! f}$,
\begin{equation}
\label{eqn:lambdaeff} \lambda_{e\! f\! f} = {\ell_p ((1-t)\lambda_a +t\lambda_b) +\ell_c
\lambda_c \over \ell_p+\ell_c}
\end{equation}
where $\ell_p$ is the in-plane crystal dimension parallel to the flow of diamagnetic
screening currents (the geometry is that of Fig.~\ref{fig:fields}). Each component of
$\lambda$ can be isolated by combining measurements performed on the {\em same crystal}
but with its geometry altered in a controlled manner as attained by rotating and cleaving
the crystal. This matter is discussed in detail in Appendix~\ref{app:lambdas}.

\section{The Crystal Field Hamiltonian}

In this section, the way in which the magnetic penetration depth is obtained from our
measurements of the ESR absorption spectrum is explained. This amounts to the treatment
of a dilute random array of Gd$^{3+}$ ions, each having an electron spin of $S= 7/2$,
that contribute to the measured quantity ${1\over 2}\mu_0\lambda\chi''(\omega)$. We point
out that in this work the small splitting due to the Gd isotopes with non-zero nuclear
spin can be ignored.\cite{comment} The spectrum is a result of the energy level
configuration determined by the splitting of the degenerate Gd spin levels by the
crystalline field. Because the spin system is dilute, it can be described by a
non-interacting single-spin effective crystal field (CF) Hamiltonian. The application of
a microwave field, oriented perpendicular to the principal spin axis of the system, is a
time dependent perturbation that induces transitions between the spin levels. When the
frequency of the applied field is tuned to the splitting between two levels, a maximum in
the ESR absorption is observed. The intensity of the transition is found through the
application of Fermi's golden rule for the two levels that define the transition, $m_i$
and $m_j$, which allows the transition's contribution to the susceptibility to be written as
\begin{equation}\label{eq:fermi}
\chi''_{ij}(\omega) = \mu_0{\pi \over
\hbar}|\mu_{ij}|^2\delta(\omega-[E_j-E_i]/\hbar)[N_i-N_j].
\end{equation}
In this expression, the eigenstates $m_i$ and $m_j$ correspond to the energies
$E_i$ and $E_j$. The level occupation number is $N_i = N_0 {\cal Z}^{-1}
\exp(-\beta E_i)$, where $N_0$ is the number of spins per unit volume, ${\cal Z}$ is the partition function and $\beta=1/k_B T$ is the
inverse temperature.  The time dependent perturbation induced by the microwave field,
 $H_{r\! f}$, is proportional to the $S_x$ or
$S_y$ operator and the matrix element between the two states is $\mu_{ij} =
g\mu_B\langle m_j | S_{x/y}|m_i \rangle$, where $g$ is the Land\'e factor and
$\mu_B$ is the Bohr magneton.

\begin{figure}
\includegraphics[width=\columnwidth]{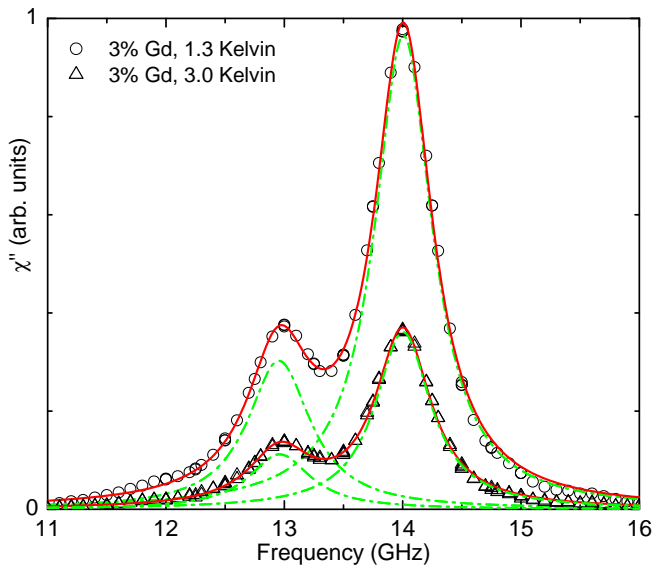}
\caption{\label{fig:temp} The temperature dependence of the ESR line width for
1.3~K and 3.0~K of an optimally doped Gd$_{\rm x}$Y$_{\rm
1-x}$Ba$_2$Cu$_3$O$_{6+{\rm y}}$ sample with x=.03, normalized by the height of
the 1.3~K peak. The G-band has line widths of 0.595 and 0.571~GHz and the
F-band has line widths of 0.649 and 0.636~GHz at 1.3~K and 3~K, respectively.
The line widths in our measurements of 1\% Gd samples presented in this work
are widths of 0.25-0.46~GHz at 1.3~K.}
\end{figure}

In real materials the ESR transitions are broadened by various relaxation processes such
as spin-spin interactions, lattice disorder and spin-lattice interactions. This
broadening, however, does not change the overall intensity of the ESR line, but merely
replaces the delta function by some line shape function (in our case it is close to
Lorentzian) whose integrated intensity is unity. In the initial stages of this work we
examined optimally doped (y=0.93) crystals having three different concentrations of Gd
ions; x$\simeq 0.5\%$, x$\simeq 1\%$, and x$\simeq 3\%$. From these measurements, we were
able to conclude that the width of the Lorentzian-shaped ESR peaks ($\delta{\rm f} =
0.35$~GHz for x$=1$\%) scaled approximately linearly with the nominal Gd concentration x,
but with a substantial x=0 intercept of about 0.30~GHz. Although we did not perform the
detailed EPMA analysis to measure x of the nominally 0.5\% and 3\% samples, it seems
clear that at x$=1$\% there is a dominant contribution to the broadening that is
concentration independent. Since a dilute spin-spin interaction is expected to provide a
relaxation rate that is linearly proportional to the concentration,\cite{AB} our results
suggest that some other process is responsible for much of the line broadening.  The line
width was also measured in the optimally doped samples at two different temperatures as
shown in Fig.~\ref{fig:temp}. The line widths decrease slightly upon increasing the
temperature from 1~K to 3~K, indicating that spin-lattice relaxation is not likely to be
the dominant process since typical relaxation mechanisms would be strongly temperature
dependent. It remains unclear what the main spin relaxation mechanism responsible for the
ESR line width is here. A final point is that no power dependence of the spectra were
ever observed verifying that saturation effects were negligible and that the assumption
of thermal equilibrium implicit in the Boltzmann factors used in Eq.~\ref{eq:fermi} is
well-founded.

For the purpose of extracting $\lambda$, the quantity of interest is the effective number
of spins exposed to the field $H_{r\! f}$.  This requires a knowledge of the energy
levels, their population and the corresponding matrix elements. This in turn requires
knowing the Hamiltonian of the system.
The Gd atom has the electron configuration of $[{\rm Xe}]4f^75d^16s^2$ and ionization
number of three in Gd$_{\rm x}$Y$_{\rm 1-x}$Ba$_2$Cu$_3$O$_{6+{\rm y}}$. This
implies that the outer shell of the ion remains exactly half filled (seven $f$
electrons in fourteen states) and the system behaves like a $7/2$ spin with no
mixing with other $S+L$ multiplets due to Hund's rule.
The construction of an effective Hamiltonian for a spin system is based on the
expansion of the crystal field around the magnetic ion in terms of Stevens
operators.  These are the operator analogues of the spherical harmonics and are
functions of $\hat S$, $\hat S_z$ and $\hat S_\pm$, given, for example, by
Abragam and Bleaney.\cite{AB} Due to the lattice symmetries and the finite spin
of the magnetic ion, the expansion reduces to a finite sum of operators $H =
\sum_{p,q}B^p_qO^p_q$ where the $B^p_q$ are numerical coefficients called the
crystal field parameters. The highest order of Stevens operator $O^p_q$ allowed
for a spin of $7/2$ in a crystal field is 6 since the matrix elements of the
Stevens operators between two spin states are zero unless $p\leq 2S$. The order
is restricted to even values by the required time reversal symmetry of the
Hamiltonian. Furthermore, the $\widehat{ab}$-plane anisotropy is so small that
for most purposes one can assume $\pi / 2$ rotation symmetry about the $\hat
c$-axis; orthorhombicity is most significant in the overdoped samples. In the
case of tetragonal symmetry, the relevant operators in order of significance
are $O^0_2,O^0_4,O^4_4$, $O^0_6$ and $O^4_6$.  The sixth order terms are
difficult to resolve in our measured spectra and we simply adopt those of
previous authors.\cite{Janossy,Pekker,Rockenbauer} When the tetragonal ordering
is slightly distorted, as will be discussed in the next section, three more
Stevens operators can contribute to the Hamiltonian.  For the present work, the
largest and only significant one of them is $O^2_2$.

\begin{figure}
\includegraphics[width = \columnwidth]{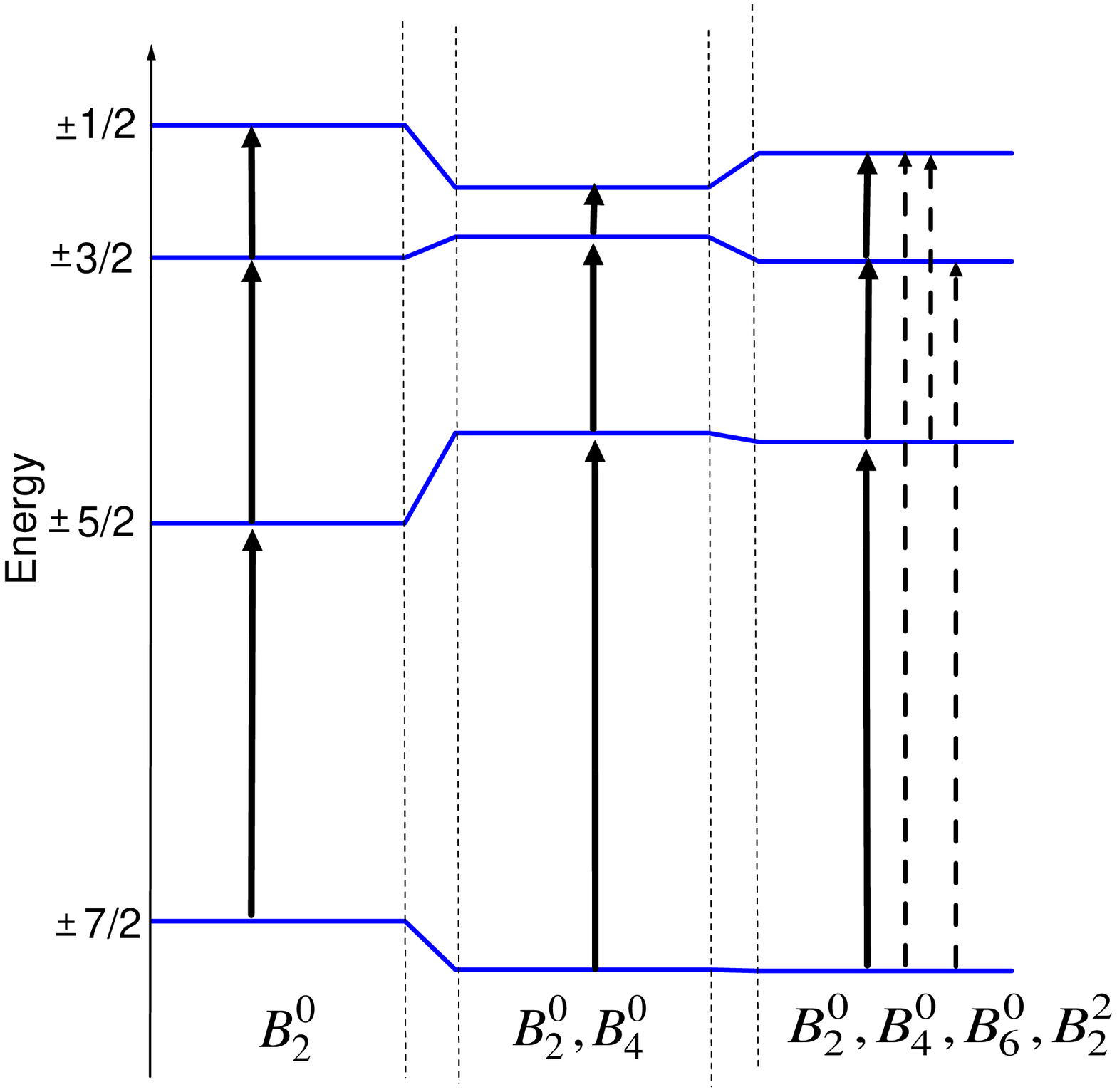}
\caption{\label{fig:levels} The energy levels resulting from the effective
crystal field Hamiltonian.  The levels, from left to right, correspond to: a
diagonal Hamiltonian, $H\propto O^0_2 = 3S_z^2-S(S+1)$, a tetragonal
Hamiltonian with some off-diagonal level mixing and finally an orthorhombic
Hamiltonian with $O^2_2 = (S_+^2+S_-^2)/2$.  The CF parameters shown here are
typical for the samples we have measured, except for $B^2_2$ which has been
exaggerated in order to make its effect more visible.}
\end{figure}

The spin $7/2$ system discussed herein has four doubly degenerate energy levels with four
eigenstates that are roughly the $\pm S_z$ eigenstates of the $S_z$ operator.  The three
allowed ESR transitions correspond to $S_z \to S_z \pm 1$. Small off-diagonal terms in
the Hamiltonian, such as the $O^2_q$ and $O^4_q$, induce small level mixing that allow
the otherwise forbidden transitions corresponding to $\pm 5/2 \to \pm 1/2$, $\pm 7/2 \to
\pm 1/2$ and $\pm 7/2 \to \pm 3/2$ to occur, albeit with relatively low intensities. The
resulting energy levels for different CF Hamiltonians are shown in Fig.~\ref{fig:levels}.
Due to the oxygen ordering in chains, which we discuss in detail in the next
section, most samples have more than one inequivalent site for the Gd ions.
Therefore, two or three sets of CF parameters are needed to fit the spectrum.
Following J\'anossy {\em et al.},\cite{Janossy,Pekker, Rockenbauer} we label
the spectra corresponding to a particular site by a band index using the
notation shown in Fig.~\ref{fig:bands}.

We begin our fitting procedure by first constructing an effective spin Hamiltonian using
the measured CF parameters reported by J\'anossy {\em et al.}~\cite{Janossy} and then
adjusting the parameters to best fit our data.  We focus on the main transition ($\pm7/2
\to \pm5/2$ for each band) because it is the strongest and consequently least susceptible
to experimental uncertainty. As a final step, a multiplicative factor representing the
effective number of spins that have participated in the ESR process is then used to scale
the overall amplitude of the model spectrum to match the measured data. This number is the effective penetration depth $\lambda_{e\! f\! f}$,
combined with the measured Gd concentration x in the crystal.

\section{Dependence of ESR spectra on oxygen configuration}

\begin{figure*}
\includegraphics[width=\textwidth]{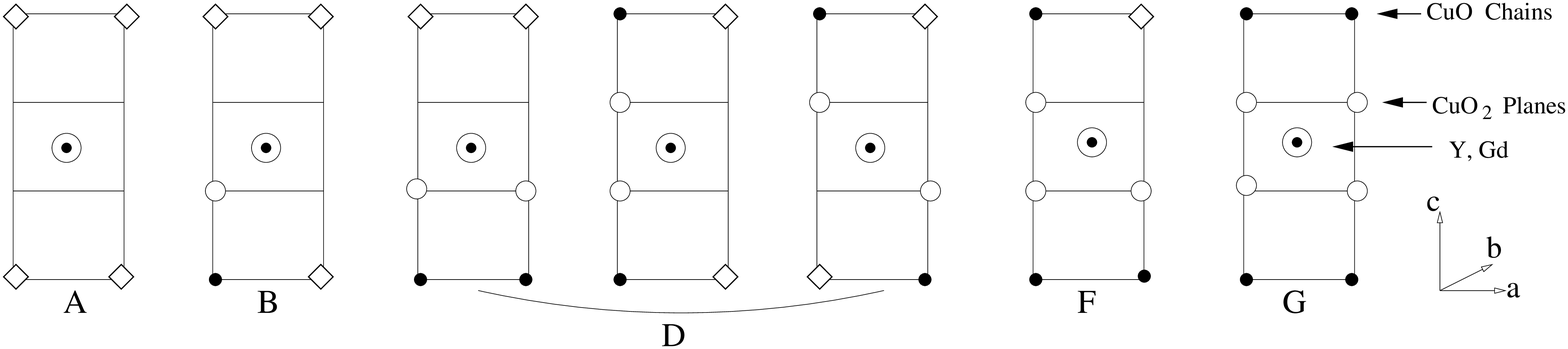}
\caption{\label{fig:bands} Identification of the ESR bands (A,B,D,F,G) attributed to a
variety of oxygen configurations, taken from S.~Pekker {\em et al.}\cite{Pekker} Figures
represent cross-sections through the center of a YBCO unit cell (with Gd located on the Y
site) parallel to the $\widehat{ac}$-plane. Open circles represent doped holes in the CuO$_2$
planes, black circles represent oxygen in the CuO chains and diamonds represent oxygen
vacancies in the CuO chains. The central circle is the Gd ion. Note that the three D-band
configurations all result in a single ESR band.}

\includegraphics[width = \textwidth]{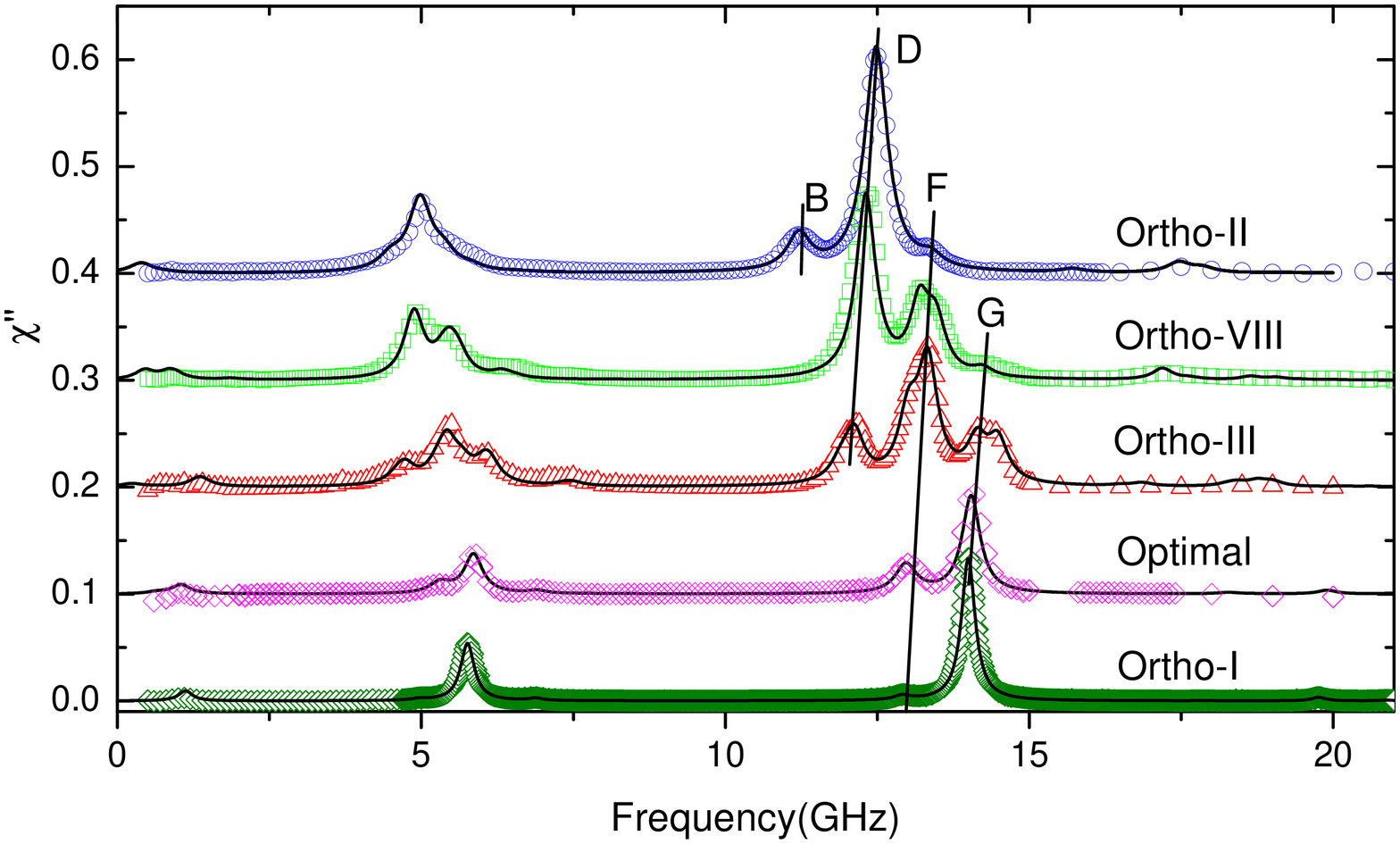}
\caption{\label{fig:chis} Measured ESR spectra for five dopings with screening currents
running along the $\hat a$ direction, shown together with the spectrum calculated using
the CF Hamiltonian. The calculated curve is scaled by an overall multiplicative factor of
$\lambda_{e\!f\!f}$ as explained in the text. The various spectra were offset by 0.1 from
each other and the ortho-I spectrum was scaled down by a factor of four.}
\end{figure*}

The presence of a chain layer breaks the tetragonal symmetry and the Stevens operator
expansion of the foregoing section is then no longer exact. However, the chains are far
from the Gd sites and this effect can be treated as a perturbation, leading to line
broadening or line splitting. Here we will consider line splitting associated with
distinct Gd environments and simply handle the line width as a parameter in a Lorentzian
fit. Different types of Gd sites are encountered because hole doping in this system is
controlled by manipulating the oxygen content and ordering of the CuO$_{\rm y}$ chains in
Gd$_{\rm x}$Y$_{\rm 1-x}$Ba$_2$Cu$_3$O$_{\rm 6+y}$. The oxygen content is set by
annealing in controlled oxygen partial pressure at high temperatures and then the oxygen
ions tend to organize into lengths of CuO chain fragments which are able to promote holes
to the CuO$_2$ planes.\cite{Zaanen} The strong tendency to form chain fragments means
that there are 7 probable Gd environments corresponding to anything from 0 to 4 nearest
neighbor chains. These different crystallographic environments have been identified in
ESR experiments on magnetically aligned Gd$_{\rm x}$Y$_{\rm 1-x}$Ba$_2$Cu$_3$O$_{\rm
6+y}$ powders\cite{Janossy,Rockenbauer,Pekker} and are illustrated in
Fig.~\ref{fig:bands}.

The chains also tend to form ordered periodic superlattices consisting of arrangements of
full and empty CuO$_{\rm y}$ chains. The structures can be particularly well-ordered for
special values of y. In the case of full oxygen doping, the CuO chains are nearly
completely full with every Gd ion having four nearest neighbor chains. This results in
the simplest of the spectra observed in the five crystals studied.  It involves one
primary set of CF parameters, denoted the G-band by J\'anossey {\em et al.}, and
generates three dominant ESR transitions. A secondary contribution of much lower
intensity is also present in these measurements, resulting from a configuration where one
of the four chains is missing oxygen, denoted the F-band. In the overdoped sample, the
largest transition ($7/2 \to 5/2$) of this latter band is barely discernable, but as the
oxygen concentration is reduced to optimal doping, the intensity of the F-band builds, as
seen in Fig.~\ref{fig:chis}.

At lower doping, the next best-ordered phase occurs near y~$\approx0.5$, where the chains
form an ortho-II structure of alternating full and empty chains.\cite{RuixingII} Other
ordered phases are: ortho-III with FFE (full, full, empty) chains and ortho-VIII with
FFEFFEFE chains. In each of these phases there is more than one possible chain
configuration around the Gd ions. Having detailed spectra at each doping allows us to
identify the different bands and fit each spectrum with a set of crystal field parameters
for each band. The fitted $\chi''$ spectra are shown in Fig.~\ref{fig:chis}. In each
spectrum, the relative intensities of the different bands is a measure of the relative
number of Gd ions in each configuration. This information can be used as a measure of the
amount of oxygen in the sample:
\begin{eqnarray}
{\rm y} = {1\over 4}\sum_n n I_n \;\;\;\;\;\; \sum_n I_n = 1
\end{eqnarray}
where $n$ is the number of full nearest neighbor CuO chains in the band and $I_n$ is the
observed relative intensity of the line associated with a particular band. The oxygen
content is given by $6+{\rm y}$. The ESR measured oxygen content versus the chemically
measured content is presented in Fig.~\ref{fig:doping}. Note that the ESR points
consistently underestimate the amount of oxygen in the crystals. This is likely due to
the inefficiency of isolated oxygens in empty chains in promoting a hole into the CuO$_2$
planes. The above oxygen content analysis assumes maximal length of chains, {\em i.e.}
all chains are either full or missing with no isolated oxygens or vacancies. Since the Gd
ions are likely to be sensing the presence of chains through their influence on the
charge distribution on the much nearer CuO$_2$ planes, an isolated oxygen ion hardly
affects the crystal field environment since it does not promote a hole into the
planes.\cite{Zaanen} Thus a Gd ion near an isolated oxygen in an otherwise empty chain
will experience the crystal field of an empty chain. This will lead to an underestimate
of the oxygen content, but will not affect the overall counting of Gd ions.

It is also interesting to note that the ESR analysis provides a good measure of the chain
disorder.  For example, perfect three dimensional ortho-II ordering should produce only one band,
corresponding to two full chains, the D-band. However, our ortho-II spectrum displays two
additional bands (the B- and F- bands) resulting from imperfections in the ortho-II
ordering.\cite{RuixingII}  These deviations can be interpreted as ortho-II phase
boundaries. Our ortho-III and ortho-VIII spectra are consistent with long range ortho-III
and ortho-VIII order along the $\widehat {ab}$ plane but random stacking along the ${\hat
c}$-direction.

\section{Results}
\begin{figure}
\includegraphics[width=\columnwidth]{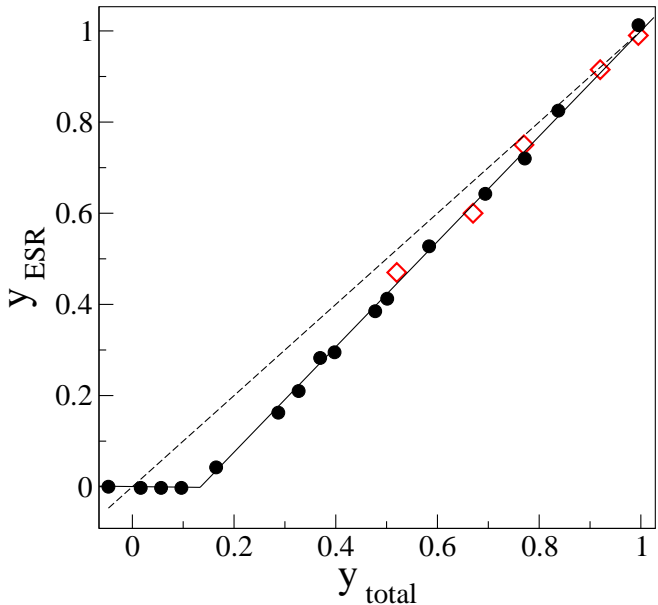}
\caption{\label{fig:doping} Ordered oxygen content as measured by ESR vs. the total
oxygen content of the sample. The solid circles are taken from Pekker~{\em et
al.}\cite{Pekker} (y$_{\rm total}$ determined by weighing the sample) and the open
diamonds are the values obtained from the present work (y$_{\rm total}$ determined by the
annealing temperature and oxygen partial pressure).}
\end{figure}
\begin{figure}
\includegraphics[width = \columnwidth]{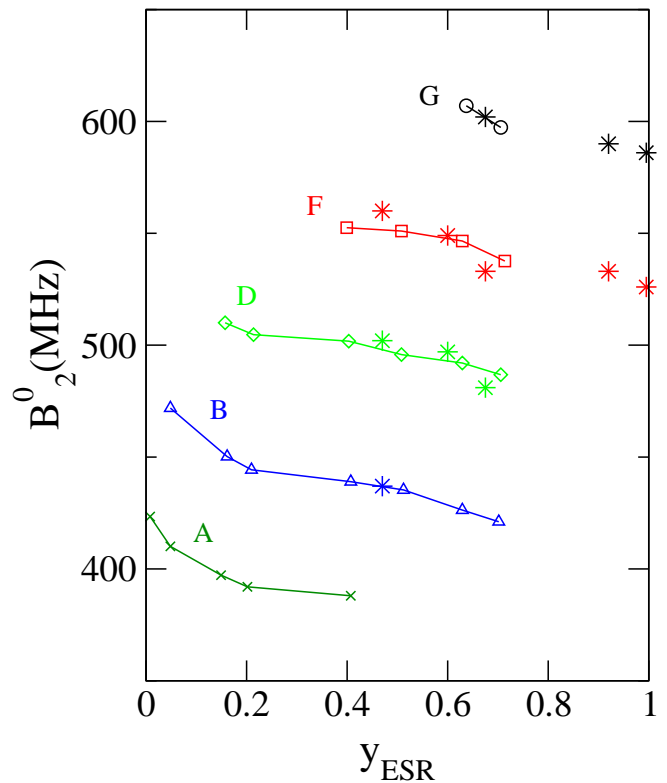}
\caption{\label{fig:CFb02} The oxygen doping evolution of the main CF parameter, $B^0_2$
in MHz for the different bands. The open symbols are taken from Rockenbauer {\em et al.}
\cite{Rockenbauer} and the stars are the ZF ESR best fit values.  For the ortho-III
sample, where more than one set of CF parameters was found for each line, we show the
values closest to previously the measured ones.}
\end{figure}
\begin{figure*}
\includegraphics[width=\textwidth]{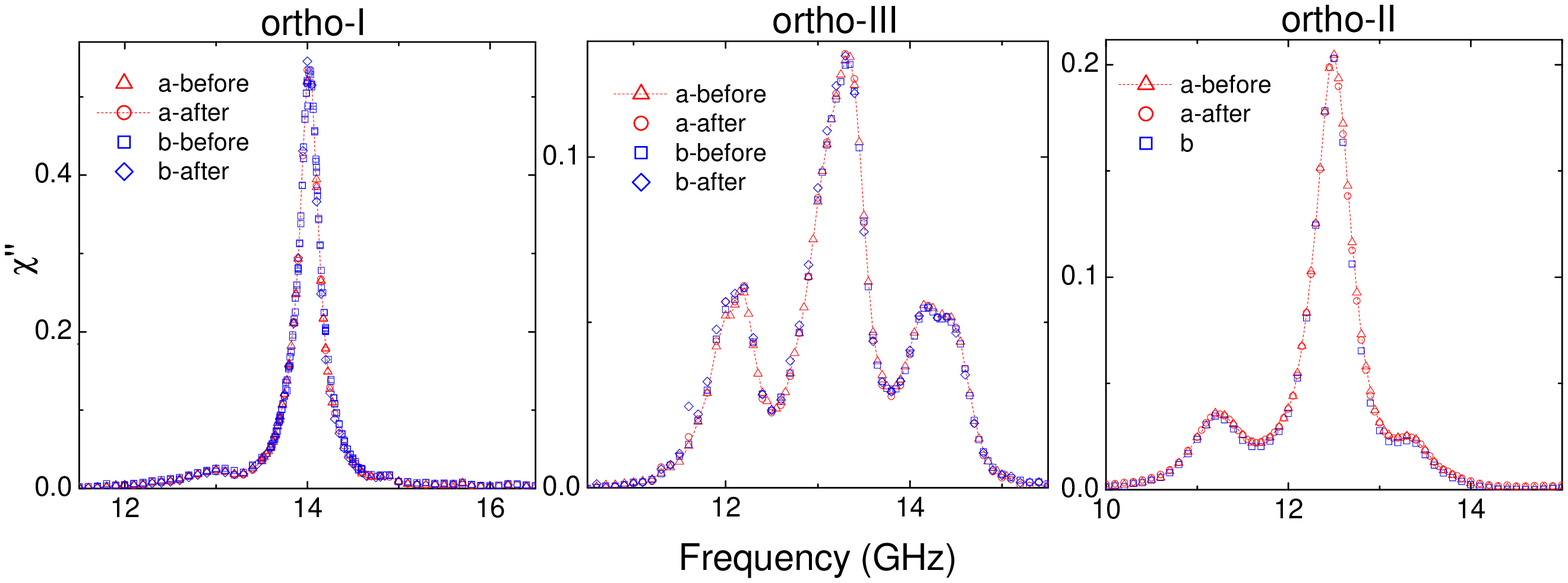}
\caption{\label{fig:cleave} The main ESR line $\pm 5/2 \to \pm 7/2$ in ortho-I,
ortho-III and ortho-II.  In each figure we overlay three or four measurements
of the same crystal in the $\hat a$ and $\hat b$-directions, before and after
cleaving the sample.  Each graph has been scaled by the effective penetration
depth $\lambda_{e\! f\! f}$ of the particular measurement so that only $\chi''$
is presented.  The changes in $\lambda_{e\! f\! f}$ are due to the different
aspect ratios of the crystal before and after the cleave and the different
orientations of measurements.}
\end{figure*}
We have performed detailed measurements of the ZF ESR absorption spectrum in high quality
samples of Gd$_{0.01}$Y$_{0.99}$Ba$_2$Cu$_3$O$_{6 + {\rm y}}$ at five different values of
y. As noted above, four of these are chosen to be in ranges with particularly
well-ordered CuO chain structures, denoted by the periodicity of their chain
superlattices (ortho-I,II,III,VIII). The optimally doped sample, with the maximum
attainable $T_c$ of 93~K (y=0.93) is also in the ortho-I phase, but has a substantial
number of oxygen vacancies on the chains. The ESR spectrum at each doping was fitted with
a spectrum generated by an effective CF Hamiltonian, with initial guesses for the CF
parameters taken from Rockenbauer {\em et al.}\cite{Rockenbauer} Subsequently, the CF
parameters were adjusted to best describe our measured spectra, and the best fit values
are presented in Table~\ref{table:CF}. The evolution of the dominant CF parameter
($B^0_2$) as a function of the ESR measured oxygen content is shown in
Fig.~\ref{fig:CFb02}. The values and systematic trends of our fit values compare well to
those of the conventional ESR studies on powders.\cite{Janossy,Rockenbauer, Pekker}

\begin{figure}
\includegraphics[width=\columnwidth]{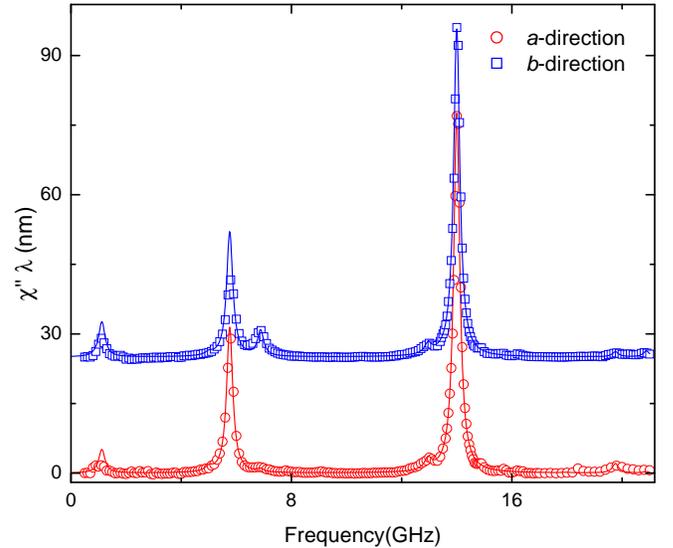}
\caption{\label{fig:oI} The oxygen-overdoped
Gd$_{.01}$Y$_{.99}$Ba$_2$Cu$_3$O$_{6.993}$ sample ESR absorption spectra in the
$\hat a$ and $\hat b$-directions. The $\hat b$-axis data were offset by 25~nm
for clarity. The orthorhombicity of the crystal lattice is apparent in the
appearance of an additional ESR line at 6.9~GHz when the screening currents
flow in the $\hat b$-direction. This line corresponds to the $\pm 1/2 \to \pm
5/2$ transition that is allowed along the $\hat b$-direction, generated by the
orthorhombic Stevens operator $B^2_2 = (S_x^2-S_y^2)/2$. This same line appears
with suppressed intensity in the $\hat a$-direction due to a small fraction of
twinning in the sample. The observed spectrum indicates a predominance of the
G-band oxygen configuration, as expected for a fully doped sample, with a 4\%
contribution from the F-band.  For comparison, note that the F-band
contribution in the optimally doped sample shown in Fig.~\ref{fig:chis} is much
larger as a result of the lower oxygen content.}
\end{figure}

The ESR spectrum is a sensitive probe of the crystallographic structure and the crystal
symmetry in particular. This high sensitivity is demonstrated at its best in the spectrum
of our overdoped sample. This is the most ordered and most orthorhombic phase, with four
full CuO chains around virtually every Gd ion. The lines are very sharp and the main
band, G, corresponds to 96\% of the total spectral weight. As seen in Fig.~\ref{fig:oI},
the orthorhombicity of the crystal is manifested as an additional line in the spectrum
that appears only when the measurement is performed with screening currents flowing along
the $\hat b$-direction. The spectra for both directions can be fit with a single set of
CF parameters by introducing an orthorhombic term in the CF Hamiltonian. This term is
$B_2^2O^2_2$, where $B_2^2$ is the measured coefficient and $O^2_2 = (S_+^2 + S_-^2)/2
=(S_x^2-S_y^2)/2$. The inclusion of this term allows the otherwise forbidden transitions
of $\pm 5/2 \to \pm 1/2$ only when the magnetic field is applied in the $\hat
a$-direction, and reveals the slight orthorhombicity of the crystal (the difference
between the $\hat a$ and $\hat b$-dimensions of the unit cell is less than 2\%).

The measured ortho-III spectrum is indicative of some disorder in the chain
structure. First, the substantial amount of spectral weight in the F-band,
which corresponds to three full chains, indicates that even if the ordering is
perfect in the planes, the stacking along the $\hat c$-direction is disordered,
{\em i.e.} that the ortho-III structure may shift between adjacent chain
layers. Second, we find that each ESR peak is composed of two closely spaced
lines that together produce a non-Lorentzian shape.  Since the measured ESR
oxygen level is consistent with the other crystals (see y$_{\rm total}= 0.77$
in Fig.~\ref{fig:doping}), we are confident in the band identification and do
not interpret the minor splitting as a change of chain configuration.  The
splitting implies a small variation of CF parameters {\it within the same
band}.  This might be the result of inhomogeneity in the oxygen concentration,
which is consistent with a measured transition temperature broadening
$\Delta$T$_c\approx$1~K. However, we do not think that the sample is
macroscopically phase separated since the line shapes were not affected by
cleaving the sample into smaller pieces.

Our ortho-II spectra suggest fairly good ordering within the chain layers with a somewhat
shorter correlation length in the $\hat c$-direction.  This is in agreement with X-ray
analysis.\cite{RuixingII} The extraction of $\lambda_c$ in this sample was done by
comparing two measurements in the $\hat a$-direction, where the first measurement was
done on the whole sample and the second one on two pieces resulting from a cleave along
the $\hat b$-direction.  The effective penetration depth, $\lambda_{e\! f\! f}$ was
observed to increase by about 210~nm due to the introduction of two more sample
$\widehat{bc}$-faces, where currents flow along the $\hat c$-direction.

We have obtained enough data in order to reliably extract the absolute values of the
penetration depth in all three crystallographic directions for three of the five doping
levels studied: fully doped, ortho-II and ortho-III. Our measurements at the remaining
two doping levels were made using samples that were not amenable to the cleaving
procedure required to extract absolute values of $\lambda$. Our final results for
$\lambda$ were derived as described in the previous sections with the emphasis on
performing enough measurements on the same sample in order to eliminate sample-dependent
effects and to overdetermine the values of $\lambda$ for error control. The penetration
depth results are presented in Table~\ref{table:lambda}.  The error estimates account for
uncertainties arising from measurements of the sample dimensions, calibration of the
microwave absorption experiment\cite{PatRSI} and the estimation of the ESR fit
parameters.  Since the extraction of the penetration depth in the three crystallographic
directions depends sensitively on the sample's dimensions, it is important to work with
crystals that have smooth parallel well defined faces whose area is easy to measure. This
was most easily achieved by choosing a single sample having a nice platelet geometry for
each doping.  If we restrict our analysis to measurements made on the same crystal where
the geometry of the sample can be altered in a controlled fashion by cleaving, we obtain
a high degree of internal consistency between different measurements. In all cases we
performed an extra measurement that overdetermines the penetration depth values and found
agreement ranging from 0.5\% to 6\%.  In the case of the ortho-II doping we measured a
second sample having a 30\% higher Gd concentration, and the results agreed to within
12\% of the first sample. The high reproducibility of our measurements is demonstrated in
Fig.~\ref{fig:cleave} where we present data for the same crystal before and after
cleaving the sample. Despite changing the aspect ratio of the crystal by approximately
50\%, the spectra are identical, modulo the overall multiplicative factor of
$\lambda_{e\! f\! f}$.

\begin{table*}\centering\label{table:CF}

\begin{tabular}{|c|c|c|c|c|c|c|c|c|c|}
\hline Sample&Band & Intensity & $B^0_2$ & $B^2_2$ & $B^0_4$ &
$B^4_4$ & $7/2 \to 5/2$& $3/2 \to 5/2$& $1/2 \to 3/2$ \\
(Gd\%)&&&(MHz)&(MHz)&(MHz)&(MHz)&(GHz)&(GHz)&(GHz)\\ \hline \hline
 ortho-I&~~G-band~~ & 96\%&-587 (-598)& 26.7 (-)& -3.1 (-3.1) & 12.5 (13.2)&14.0 &5.8 & 1.12~ \\
 \cline{2-10}
$(1.08\pm0.06)$&F-band & 4\% &-526 (-541)& 26.7 (-)& -3.1 (-3.1) &  12.5 (17.2)&12.9 &5.0 & 0.9 \\
\hline \hline
 optimal & G-band & 75\% &-590(-598)& 0& -3.1 (-3.1) & 12.7 (13.2)
&14.04 & 5.86&1.05 \\ \cline{2-10} $(1.28\pm0.02)$ &F-band & 25\% &-533(-541)& 0& -3.1
(-3.1) &  12.5 (17.2) &13.02 &5.2&  \\ \hline \hline
 ortho-III & G-band & 21\% &-617,-604,-585(-598)& 30 (-)& -3.0 (-3.1) & 12.5 (13.1)
&14.4,14.1 & 6.13,6.03&1.4,1.3 \\  \cline{2-10}
 $(1.02\pm0.05)$&F-band & 56\% &-533,-517 (-541)& 6.7 (-)& -3.5 (-3.1) &  17.1 (17.2) &13.3,13.0
&5.4,5.45&0.2,0.3  \\  \cline{2-10}
&D-band & 23\% &-481,-475 (-497)& 26.7 (-)& -3.1 (-3.1) &  14.6 (14.7) &12.1,11.9 &4.7,4.6&0.6 \\
\hline \hline ortho-VIII & G-band & 2\% &-630 (-598)& 1.6 (-)& -2.5 (-3.1) &  9 (13.1)
&14.2 & 6.3&1.9 \\  \cline{2-10} $(1.38\pm0.1)$ &F-band & 36\% &-549 (-541)& 1.6 (-)&
-3.0 (-3.1) &  12.5 (17.2) &13.2 &5.5&0.8  \\  \cline{2-10}
&D-band & 62\% &-497 (-497)& 1.6 (-)& -3.1 (-3.1) &  12.5 (14.7) &12.3 &4.9&0.8 \\
\hline \hline ortho-II&F-band & 4\% &-560 (-541)& 5 (-)& -3.1 (-3.1) &  17.2
(17.2)&13.4&5.8&0.8\\ \cline{2-10} $(1.08\pm0.06)$ &D-band & 81\% &-502 (-497)& 5 (-)&
-3.2 (-3.1) &  14.7 (14.7)&12.5&5.0&0.4\\  \cline{2-10}
 &B-band& 15\% &-437 (-445)& 5 (-)& -3.2 (-3.1) &  15.5
(17.2)&11.2&4.5&0.2 \\ \hline
\end{tabular}
\caption{Crystal field parameters for the various ESR bands for different oxygen contents
of Gd$_{.01}$Y$_{.99}$Ba$_2$Cu$_3$O$_{\rm 6+y}$, given in MHz. The Gd\% in brackets
indicates the relative substitution of Gd for Y. The CF parameters measured by
Rockenbauer {\it et al.}\cite{Rockenbauer} in conventional ESR on powder samples are
given in brackets, and at each doping we have taken values from the nearest doping
measured by these authors. The values of $B^0_6$ and $B^4_6$ could not be resolved from
our spectra and we simply adopt the values of Rockenbauer {\it et al.} The transition
frequencies, shown in GHz, were found either directly from the observed lines or by the
fit Hamiltonian (when the lines were unobservable). The ortho-III spectrum required more
than one set of CF parameters for each band due to further line splitting of unknown
origin, which may be related to oxygen inhomogeneity.}
\end{table*}

\begin{table*}\centering
\begin{tabular}{|c|c|c|c|c|c|c|c|}
\hline ~Crystal Structure~ &~~~~~$T_c$(K)~~~~& ~Oxygen Content~ & ~~~~$\lambda_a$(nm)~~~~
& ~~~~$\lambda_b$(nm)~~~~ & $\lambda_a/\lambda_b$& ~~~~$\lambda_c$(nm)~~~~  \\ \hline
\hline Ortho-I & 89&6.995 & $103\pm 8$ & $80\pm 5$& $1.29 \pm 0.07$& $635\pm 50$\\
\hline Ortho-III &75 &6.77 &$135\pm 13$ & $116\pm 12$ & $1.16 \pm 0.12$& $2068\pm 200$\\
\hline
Ortho-II &56& 6.52  &$202\pm 22$ & $140\pm 28$ & $1.44 \pm 0.26$ & $7500 \pm 480$ \\
\hline
\end{tabular}
\caption{Experimental values of the anisotropic magnetic penetration depth extracted from
four or more different measurements of $\lambda_{e\! f\! f}$ on the same crystal, with
the microwave field applied in different directions and for different sample aspect
ratios achieved by cleaving. The uncertainty in the ansisotropy is reduced since some of
the systematic contributions to the uncertainty in absolute value cancel when comparing
measurements on the same sample.}

\label{table:lambda}

\end{table*}

\section{discussion}

The results presented here have impact in two areas. First, we have provided detailed
information about the Gd$_{\rm x}$Y$_{\rm 1-x}$Ba$_2$Cu$_3$O$_{\rm 6+y}$ system, namely
the crystal field parameters at the yttrium site and the way in which oxygen orders into
CuO chains. We have shown that our crystal field parameters agree very well with those
obtained by previous authors using the completely different method of high field ESR on
powdered samples. Our results also support the conclusions of the same work which found
that isolated oxygens in vacant chains do not influence the CF configuration around the
magnetic ion and therefore cannot be distinguished from a vacant chain via ESR
spectroscopy. This is in accord with the fact that isolated oxygens do not promote holes
from the copper-oxygen planes.\cite{Zaanen}

The principal result of this work is a set of new values for the penetration depth in all
three crystallographic directions for three oxygen-ordered phases of YBa$_2$Cu$_3$O$_{\rm
6+y}$ which are summarized in Table II. The $\hat{c}$-axis penetration depth is large and
increases very rapidly with decreasing doping, as observed previously in infrared
measurements of $\lambda_c$ by Homes {\em et al.}\cite{Homes} The in-plane measurements
are in accord with the results of Sonier {\em et al.}\cite{Sonier1,Sonier2} who performed
$\mu SR$ measurements on mosaics of high-purity, detwinned single crystals at two other
oxygen dopings. The muon measurements are unable to directly determine anisotropies, but
they compare well with the geometric mean of the $\widehat{ab}$-plane values reported here. A
sample of YBa$_2$Cu$_3$O$_{6.60}$ with $T_c$ = 59 K was found to have $\lambda_{ab}$ =
170 nm,\cite{Sonier2} very close to the geometric mean $\sqrt{\lambda_a\lambda_b}$ =
168$\pm$26 nm of the $T_c$ = 56 K sample studied here. An optimally doped sample of
YBa$_2$Cu$_3$O$_{6.95}$ with $T_c$ = 93 K had $\lambda_{ab}$ = 112 nm,\cite{Sonier1}
which lies between the means of the $T_c$ = 75 K sample (125$\pm$12 nm) and the overdoped
$T_c$ = 89 K sample (91$\pm$6 nm) studied here. The agreement reflects the particular
care taken in these $\mu SR$ measurements to cover a wide range of applied magnetic
fields in order to ascertain the low field limiting values and thus minimize the
non-linear, non-local, and other effects that can arise in the vortex state. It must also be
noted that the $\mu SR$ measurements hinge on a detailed model of the vortex lattice that
provides a fit to the field distribution detected by the muons.

These absolute values of penetration depth obtained in single crystals are smaller than
many measurements found in the existing literature on YBa$_2$Cu$_3$O$_{6+y}$.
Magnetization studies of aligned powders by Panagopoulos {\em et al.}\cite{Panagopoulos}
gave 140 nm for $T_c$=92, 210 nm for $T_c$=66 and 280 nm for $T_c$=56 (all $\pm$25\%),
which are longer than the values presented here. The origin of this discrepancy is not
clear, but might be due to assumptions made in the analysis of the powder data or
problems with the surfaces of grains embedded in epoxy. Our absolute values of
$\lambda_a$ and $\lambda_b$ are also smaller than those obtained by far infrared
measurements near optimal doping\cite{Basov} ($\lambda_a$=160 nm, $\lambda_b$=100nm).
However, the infrared measurements did point out the importance of the in-plane
anisotropy in these materials, an anisotropy that becomes very large in the
YBa$_2$Cu$_3$O$_{6.99}$ sample studied here. This anisotropy has been attributed to the
presence of a nearly one-dimensional Fermi sheet derived mainly from the bands associated
with the CuO chains.\cite{Atkinson} Although $\mu SR$ measurements cannot directly
measure this anisotropy, Tallon {\em et al.}\cite{Tallon} also inferred such a
contribution by noting the very large increase in muon depolarization rates as the chain
oxygen sites become filled near YBa$_2$Cu$_3$O$_{7}$. The interpretation of the
polycrystalline data suggested values of $\lambda_a$ = 155 nm and $\lambda_b$ = 80 nm for
samples with fully doped CuO chains, which overestimates both the overall magnitude and
the anisotropy of the in-plane penetration depth. However, the basic picture of
chain-driven anisotropy is supported by our new microwave measurements, which also show
substantial anisotropy in the ortho-II ordered sample, but rather less in the ortho-III
sample, a sensible trend since the ortho-III ordering is much poorer, leading to more
fragmented chains.

Much of the work on the doping-dependence of $\lambda_{ab}$ has relied on muon spin
relaxation measurements on polycrystalline samples. In these measurements, the field
inhomogeneity in the vortex state gives rise to dephasing of the precessing spins of
implanted muons. A Gaussian fit is often used to extract a relaxation rate $\sigma$ which
is deemed proportional to $1/\lambda^2$. As long as the material is nearly two
dimensional, the constant of proportionality can be calculated to be $\sigma = 7.09\times
10^4 \lambda^{-2}$, with $\sigma$ in $\mu s^{-1}$ and $\lambda$ in
nm.\cite{Bernhard,Bardford} In the YBa$_2$Cu$_3$O$_{\rm 6+y}$ system, this very simple
treatment of muon data gives values of $\lambda_{ab}$ that are typically 20\% or more
larger than the single crystal data reported
here.\cite{Bernhard,Uemura,Tallon2003,Pumpin} The resulting underestimate of the
superfluid density, together with the problematic $\widehat{ab}$-plane anisotropy, mean
that the single crystal penetration depths reported here can offer a clearer picture of
the doping dependence of the superfluid density in the CuO$_2$ planes.

\begin{figure}
\includegraphics[width=\columnwidth]{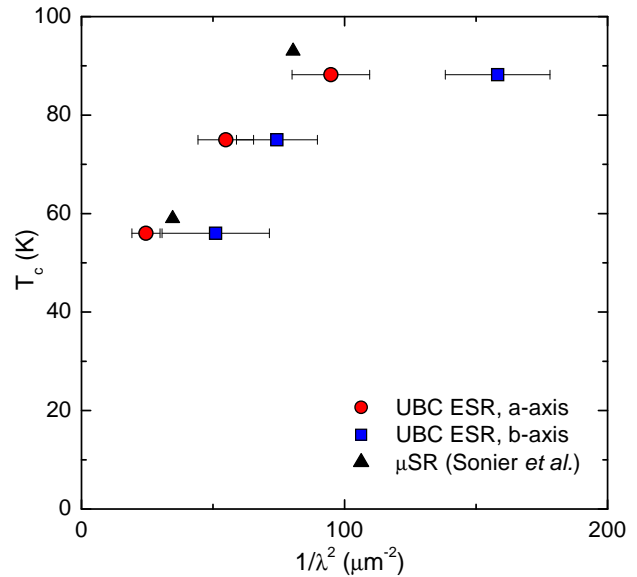}
\caption{\label{fig:Uemura} Measurements of the in-plane anisotropic absolute penetration
depth $\lambda$ in YBa$_2$Cu$_3$O$_{\rm 6+y}$ for y=0.52, y=0.77 and y=0.995 plotted as
$T_c$ versus superfluid phase stiffness ($\propto 1 / \lambda^2$). Also shown are the
$\mu SR$ measurements of Sonier {\em et al.} on single crystal mosaics.  The data do not
support a case for scaling of $T_c \propto 1/\lambda_a^2$ in this doping range above
$T_c$ = 56~K.}
\end{figure}

Figure~\ref{fig:Uemura} displays $T_c$ versus $1/\lambda^2$, a plot first suggested by
Uemura {\em et al.}\cite{Uemura} A long-accepted result has been a linear relationship
between these two quantities at low doping, followed by a plateau at higher doping that
varies from one cuprate system to another, inferred mainly from $\mu SR$ measurements on
ceramics.\cite{Uemura} The data in Fig.~\ref{fig:Uemura} confirms the overall feature
that $1/\lambda^2$ increases with $T_c$. However, the linear regime at low doping, which
implied a superfluid phase stiffness proportional to $T_c$ on the underdoped side of the
cuprate phase diagram, is not supported by the new data presented here, at least for $T_c
\ge$ 56 K. The measurements on $\rm YBa_2Cu_3O_{6.993}$, which is slightly past optimal
doping, are certainly up in the ``plateau" regime for this material and would not be
expected to follow $T_c \propto 1/\lambda^2$. However, even for the lower two dopings
with $T_c$ = 56 K and $T_c$ = 75 K the $\hat a$-axis phase stiffness, which avoids
contributions associated with the CuO chains, is still far from a relationship of the
form $T_c\propto 1/\lambda_a^2$. The plot instead suggests a sublinear dependence of
$T_c$ on $1/\lambda_a^2$, since $T_c$ must fall to zero when $1/\lambda^2$ does. Further
data at low doping will be needed to clarify the relationship over the entire doping
range. Interestingly, the latest muon spin relaxation studies by Tallon {\em et
al.}\cite{Tallon2003} also suggest a sublinear relationship, although the values of
$\lambda$ in that study are significantly larger than the single crystal data presented
here. In the past, $T_c\propto 1/\lambda^2$ has been seen as evidence that phase
fluctuations play a central role in setting $T_c$ on the underdoped side of the cuprate
phase diagram.\cite{Kivelson} Indeed, the relatively low values of phase stiffness in the
cuprates, plus the observation of critical fluctuations near the superconducting
transistion,\cite{Kamal,Meingast} do indicate that phase fluctuations play a role in
determining $T_c$. However, the deviation from a linear relationship seen in
Fig.~\ref{fig:Uemura} suggests that other factors must also contribute to the critical
temperature. The obvious candidate is thermal excitation of nodal
quasiparticles,\cite{Scalapino,Carbotte,Lee} which give rise to the linear temperature
dependence of $\lambda(T)$ and rapidly deplete the superfluid density as temperature
rises. However, the central puzzle regarding the correlation between superfluid density
and $T_c$ still remains in such a scenario. As the doping decreases, the zero temperature
value of the superfluid density becomes much smaller, so that it becomes easy for
quasiparticle excitations, in addition to fluctuations, to drive the material normal at a
lower $T_c$. The mystery still lies in understanding why the superfluid density becomes
so small with decreasing hole doping.

\acknowledgments

The authors are grateful to P.~Dosanjh for his technical assistance, and are particularly
indebted to A.~J\'anossy for sharing his knowledge of Gd ESR and providing figures
\ref{fig:doping} and \ref{fig:CFb02}. We are also grateful for useful discussions with
D.~M.~Broun, J.~Sonier, I.~Herbut, K.~Moler, J.~Guikema and S.~Kivelson. This work was
supported by the Natural Science and Engineering Research Council of Canada and the
Canadian Institute for Advanced Research.

\appendix

\section{Extracting anisotropic $\lambda$ values from multiple measurements}
\label{app:lambdas}

YBa$_2$Cu$_3$O$_{\rm 6+y}$ single crystals have strong natural cleave planes
perpendicular to both the [100] and [010] directions.  Using this fact,
previous studies of the temperature dependence of $\lambda$ have been
successful at separating the out-of-plane response from the in-plane response
by measuring a sample, cleaving it $N$ times, and then repeating the
measurement with all of the pieces together. The second measurement contains
the same in-plane contribution, but has the $\hat{c}$-axis contribution
enhanced by a factor of $N$.\cite{Ahmad} For practical reasons, in the present
work it was not possible to measure all of the crystal fragments together
following cleaving in all cases, instead we successively measured each crystal
after reducing the in-plane area. The $\hat{c}$-axis thickness $\ell_c$ of the
crystal was unchanged for all measurements, but the measurement of $\ell_c$
retains the largest relative uncertainty because $\ell_c \ll \ell_{ab}$. Since
$\lambda_c$ is also rather large, care must be taken when combining
measurements in order to minimize the impact of the large relative uncertainty
in the quantity $\ell_c \lambda_c$ in Eq.~\ref{eqn:lambdaeff}. The method we
use here is to combine measurements of the {\em same crystal} in both
$\hat{a}$-axis and $\hat{b}$-axis directions before and after cleaving where
the crystal's aspect ratio has changed by approximately 50\%, providing an
overdetermination of the three unknown $\lambda$ values. For example, before
and after cleaving the sample we measure
\begin{eqnarray}
\nonumber \lambda_{e\! f\! f}^a &=& {a((1-t)\lambda_a + t\lambda_b)+c\lambda_c \over a+c}
\\ \nonumber \lambda_{e\! f\! f}^b &=& {b(t\lambda_a + (1-t)\lambda_b)+c\lambda_c \over b+c}
\end{eqnarray}
which sum to give
\begin{eqnarray}
\label{eqn:lambdasolve} {a+c \over a}\lambda_{e\! f\! f}^a+{b+c \over b}\lambda_{e\! f\!
f}^b = \lambda_a+\lambda_b + \left( {c\over a}+{c\over b}\right)\lambda_c.
\end{eqnarray}
A second measurement with new dimensions $\ell_a^\prime$ and $\ell_b^\prime$
provide a second expression having the same form as Eq.~\ref{eqn:lambdasolve},
and subtracting the two eliminates $\lambda_a$ and $\lambda_b$, leaving only
$\lambda_c$ in terms of measured quantities. The values of $\lambda_a$ and
$\lambda_b$ are then calculated from the above expressions.

\end{document}